\documentclass[12pt]{article}
\pdfoutput=1
\usepackage{graphics, color}
\usepackage{graphicx}
\usepackage{amsmath}
\usepackage{amssymb}
\usepackage{amsfonts}

\newcommand{\sect}[1]{\section{#1}\setcounter{equation}{0}}

\def\gsim{\, \rlap{$>$}{\lower 1.1ex\hbox{$\sim$}}\,}
\def\lsim{\, \rlap{$<$}{\lower 1.1ex\hbox{$\sim$}}\,}
\newcommand{\be}{\begin{equation}}
\newcommand{\ee}{\end{equation}}
\newcommand{\bea}{\begin{eqnarray}}
\newcommand{\eea}{\end{eqnarray}}

\begin{document}


\begin{titlepage}
\bigskip
\bigskip\bigskip\bigskip
\centerline{\Large \bf Dualities of Fields and Strings}

\bigskip\bigskip\bigskip
\bigskip\bigskip\bigskip

 \centerline{
 {\bf Joseph Polchinski}}
 \bigskip
\centerline{\em Department of Physics}
\centerline{\em University of California}
\centerline{\em Santa Barbara, CA 93106 USA}
\bigskip
\centerline{\em Kavli Institute for Theoretical Physics}
\centerline{\em University of California}
\centerline{\em Santa Barbara, CA 93106-4030 USA}
\bigskip
\centerline{\tt joep@kitp.ucsb.edu }
\bigskip\bigskip\bigskip

\begin{abstract}
Duality, the equivalence between seemingly distinct quantum systems, is a curious property that has been known for at least three quarters of a century.  In the past two decades it has played a central role in mapping out the structure of theoretical physics.  I discuss the unexpected connections that have been revealed among quantum field theories and string theories.  Written for a special issue of Studies in History and Philosophy of Modern Physics.
\end{abstract}
\end{titlepage}

\baselineskip = 16pt
\tableofcontents

\baselineskip = 18pt

\setcounter{footnote}{0}

\sect{Introduction}

Perturbation theory is a central part of the education of a physicist.  One first learns the basic solvable systems, most notably the simple harmonic oscillator.  One then learns how to approach general problems in which the Hamiltonian is of the form
\be
H = H_0 + g H_{1} \,, \label{ham}
\ee
where $H_0$ is solvable and the parameter $g$ is small.  These approximation schemes give physical quantities as a perturbation series,
\be
{\cal A} = {\cal A} _0 + g {\cal A} _1 + g^2 {\cal A} _2 + \ldots\,.  \label{perts}
\ee
Here ${\cal A} $ might be an energy level, a scattering amplitude, or any other quantity of interest.\footnote{To avoid confusion, it should be noted that in some cases only even powers of $g$ appear, depending on the structure of the Hamiltonian.  Also, the series for a given physical quantity may begin with a term of order $g^m$ with nonzero $m$, rather than $m=0$ as is written here for simplicity.}  In particular, a major focus of the standard quantum field theory (QFT) course is the development of the series~(\ref{perts}) in terms of Feynman graphs.

The series~(\ref{perts}) is known not to converge in most systems of interest, in particular quantum field theories~\cite{Dyson:1952tj}.  Nevertheless, it is valuable as an asymptotic series, meaning that for $g$ sufficiently small a few terms give accurate results.  For example, in quantum electrodynamics, where the effective expansion parameter is $\alpha/2\pi \sim 10^{-3}$, this has allowed the magnetic moment of the electron to be calculated to one part in $10^{12}$.
However, as $g$ increases, perturbation theory becomes increasingly inaccurate, and it can completely miss important qualitative effects.   In the Standard Model, quark confinement is the  most notable example of such a nonperturbative effect, but others include the spontaneous breaking of chiral symmetry by condensation of quarks, and the violation of baryon and lepton numbers by instantons and skymions in the weak interaction.

There are no general methods for studying QFT's at large $g$.  In principle, the nonperturbative definition of QFT by means of the path integral plus the renormalization group, as given by Wilson~\cite{Wilson:1993dy}, implies that any physical quantity can be calculated on a large enough computer.  In practice, the theories and observables for which this can be done are limited.  Another tool in the study of QFT's is the limit of a large number of fields~\cite{Stanley:1968gx,'tHooft:1973jz}.  Here the graphical expansion simplifies and in some cases can be summed, giving a description of physical phenomena that cannot be seen in the individual terms of the series.  This is most successful for theories where the many fields organize into a vector $\phi_i$.  For matrix fields $\phi_{ij}$, including the important case of Yang-Mills fields, the graphical expansion simplifies enough to allow interesting general conclusions, but usually there are still too many graphs to sum explicitly.

A new tool, which has risen to prominence in the last two decades, is weak/strong duality, also known as $S$-duality.  In some cases it is possible to decompose the Hamiltonian in multiple ways,
\bea
H &= &H_0 + g H_{1}  \nonumber\\\
&=& H'_0 + g' H'_1 \,. \label{dualh}
\eea
Now one has two perturbative expansions.  The original $H_0$ will have a simple expression in terms of some fields $\phi$, while $H'_0$ will have a simple expression in terms of some new set of fields $\phi'$.  The $\phi'$ are related to $\phi$ in a complicated and usually nonlocal way; we will see examples in \S2.  
Typically the couplings $g$ and $g'$ have a relation of a form something like
\be
g' = 1/g\,.  \label{ggp}
\ee  
If this is so, then as $g$ becomes very large, $g'$ becomes very small, and the perturbation series in $g'$ becomes an accurate description of the system just where the series in $g$ becomes useless.  Of course, if $g \approx 1 \approx g'$ then neither expansion gives a good {\it quantitative} description, but having an understanding of the two limits gives a powerful global picture of the physics.  Phenomena that are  complicated in one description are often simple in the other.  In many interesting systems there are multiple coupling constants, and multiple dual representions~(\ref{dualh}).  In \S2.5 we will give an example where there are two coupling constants and an {\it infinite} number of dual descriptions. 

There is another, perhaps deeper, way to think about the duality~(\ref{dualh}).  In quantum field theories, the expansion in $g$ is essentially the same as the expansion in $\hbar$.  To see this, consider Yang-Mills theory, whose field strength is 
\be
\hat F_{\mu\nu} = \partial_\mu \hat A_\nu -  \partial_\nu  \hat A_\mu + g [ \hat A_\mu ,  \hat A_\nu] \,.
\ee
The Yang-Mills connection $\hat A_\mu$ is written here as a matrix.  It is useful to work with a rescaled field $ A_\mu = g \hat   A_\mu$, so that
\be
F_{\mu\nu} \equiv \left (\partial_\mu  A_\nu -  \partial_\nu  A_\mu + [ A_\mu ,  A_\nu] \right) = g \hat F_{\mu\nu} \,.  \label{gscale}
\ee
The Yang-Mills action is then
\bea
S_{\rm YM} &=& -\frac{1}{2} \int d^4x\, {\rm Tr}(\hat  F_{\mu\nu}  \hat  F^{\mu\nu}) \nonumber\\
&=& -\frac{1}{2g^2}  \int d^4x\, {\rm Tr}( F_{\mu\nu} F^{\mu\nu} )\,,
\eea
where the matrix trace makes the action gauge-invariant.
In the latter form the coupling $g$ appears only as an overall factor in the action.  Quantum amplitudes are obtained from the path integral
\be
\int {\cal D}  A \, e^{iS_{\rm YM}/\hbar} = \int {\cal D}  A \, e^{i \int d^4x\,{\rm Tr}( F_{\mu\nu}  F^{\mu\nu})/2 g^2 \hbar} \,. \label{zpath}
\ee
Note that the parameters $g$ and $\hbar$ appear only in the combination $g^2 \hbar$, so that the perturbation series for typical observables is 
\be
{\cal A} = {\cal A} _0 +( g^2 \hbar) {\cal A} _2 + (g^2  \hbar)^2 {\cal A} _4 + \ldots\,.  \label{zpert}
\ee

It follows that the small-$g$ and small-$\hbar$ limits are the same: weak coupling corresponds to the classical field limit.  When $g^2 \hbar$ is small, the exponent in the path integral~(\ref{zpath}) is large and so the integral is highly peaked on  configurations where $S_{\rm YM}$ is stationary; these are the solutions to the classical equations of motion.  When $g^2 \hbar$ is large, the path integral is not very peaked and the quantum fluctations are large.  However, when the duality~(\ref{dualh}) holds, we can change to the primed description, and now the expansion parameter is $g'^2 \hbar = \hbar/g^2$, giving a highly peaked action.  Essentially what is happening is that in the original description the fields $\phi$ have wild quantum fluctuations at large $g$, but we can find new fields $\phi'$ which behave classically.  This is a bit like a Fourier transform, where a function that is narrow in $x$ is wide in $p$, and vice-versa; we will make this analogy more precise in \S2.3. (Having made this point, we will now revert to the quantum field theorist's conventions $\hbar = c = 1$.)

The interpretation of a duality is then that we have a single quantum system that has two classical limits.  Quantum mechanics is sometimes presented as a naive one-to-one correspondence between classical and quantum theories.  In this view we quantize the classical theory to go in one direction, and take the classical limit to go in other.  Of course there are exceptions; for example, a classical gauge theory with anomalies cannot be consistently quantized.  But with dualities, a single quantum theory may have two or more classical limits.  `Quantizing' any of these produces the same quantum theory.

The original wave-particle duality already exemplified this idea: given a QFT, one can take two different classical limits depending on what one holds fixed.  One limit gives classical fields, the other classical particles.  It is fruitless to argue whether the fundamental entities are particles or fields.  The fundamental description (at least to the extent that we now understand) is a \mbox{QFT}.  Similarly it is fruitless to argue whether $\phi$ or $\phi'$ provide the fundamental description of the world; rather, it is the full quantum theory. 

With the dualities~(\ref{dualh}), the functional forms of $H_0$ and $H_1$ in terms of $\phi$ may be the same as those of $H_0'$ and $H'_1$ in terms of $\phi'$.  In this case we would say that the theory is self-dual.  Alternatively, the functional forms and even the nature of the fields may be quite different: in this case we have two very different ways to think about the system.  The term $S$-duality is applied in either case; in some cases self-duality may be implied by the context.

It is not clear why the structure of theoretical physics is so kind to us, in providing simple description in many limits that would seem to be very complicated.  It may be that there are fewer consistent quantum theories than classical ones, so that we necessarily  get to the same quantum theory from multiple classical starting points.

In \S2 we discuss dualities where both descriptions are QFT's.  We begin with some classic examples, namely the Ising model, bosonization, and free electromagnetism, where the duality can be constructed rather explicitly.  We then move on to richer examples, in particular supersymmetric Yang-Mills theories.  For these, the duality is not proven but inferred.  We discuss the evidence and the logic that supports the existence of these dualities.  We also discuss the role of supersymmetry.

In \S3 we discuss dualities between string theories.  We begin with $T$-duality, which connects two weakly coupled string theories and can be demonstrated rather explicitly.  It illustrates a number of remarkable features of string theory: that space is not fundamental but emergent, and that strings perceive spacetime geometry in a rather different way from pointlike particles and fields.  We then discuss weak/strong dualities in string theory, and the significance of branes.  A notable conclusion is that there is only a single quantum theory in the end: what appear to be different string theories are different classical limits of a single quantum theory, whose full form is not yet known.  The same analysis reveals the existence of new classical limits, which are not string theories at all. 

In \S4 we discuss dualities in which one description is a QFT and the other a string theory.  The existence of such dualities is remarkable, because QFT's are well-understood conceptually, while string theories include quantum gravity and so present many conceptual puzzles.  In fact, field-string duality currently plays a key role in providing a precise definition of the quantized theory of strings and gravity.  We describe how two puzzles, the black hole entropy and the black hole information paradox, have been clarified by dualities, although important questions remain open.  We also discuss the holographic principle, in which the emergent nature of spacetime is even more radical.   We conclude by discussing various open questions.

Apology: this is a rather sweeping subject, and I certainly have not set out to reconstruct the entire history of its development.  I have tried to choose references that will be useful to the intended audience.

\sect{QFT/QFT dualities}
\setcounter{equation}{0}

\subsection{The Ising model}

A simple example of duality appears in the two-dimensional Ising model.  This model is given by spins $\sigma_i$ living at the sites of a two-dimensional square lattice.  The subscript $i$ labels the lattice sites, and each spin can take the values $\pm 1$.  The action is
\be
S_{\rm Ising} = - \sum_{\rm links} K \sigma_i \sigma_j \,.
\ee
The sum runs over all nearest-neighbor pairs, i.e.\ links, of the lattice.  The path integral is\footnote{To be precise, this is the analog of the Lagrangian form~(\ref{zpath}), in a Euclidean spacetime.  The Hamiltonian form would have a discrete spatial direction but continuous time, which could either be Euclidean or Lorentzian.  The duality exists in these cases as well.}
\be
 \left(\prod_i \sum_{\sigma_i = \pm 1}\right) e^{-S_{\rm Ising}} \,.
\ee
The parameter $K$ plays the role of $1/g^2  \hbar$.  When it is large, the path integral is highly peaked on the configurations that minimize the action, namely all $\sigma_i$ being equal (either +1 or $-1$).  When it is small, the path integral receives contributions from many configurations.\footnote{For the Ising model, a strong coupling (small $K$) expansion actually exists: the discreteness of the spacetime and the boundedness of the spins allow an expansion in powers of $K$.  Moreover, this system is exactly solvable for all $K$~\cite{Onsager:1943jn}.  We will not make use of these special  properties; our purpose is to illustrate weak/strong duality, which is more general.}  

Kramers and Wannier~\cite{Kramers:1941kn} showed that the path integral could also be written as
\be
\left(\prod_i \sum_{\sigma'_i = \pm 1}\right) e^{-S_{\rm Ising}'} \,, 
\ee
where
\be
S'_{\rm Ising} = - \sum_{\rm links} K'  \sigma'_i \sigma'_j \,, \quad
K' = -\frac12 \ln \tanh K \,,\quad K = -\frac12 \ln \tanh K' \,. \label{isingdual}
\ee
The new variables $\sigma'_i$ also live on a square lattice and take the values $\pm 1$.  The relationship of $K$ to $K'$ is more complicated than the simple reciprocal~(\ref{ggp}), but shares the property that $K' \to \infty$ as $K \to 0$ and vice versa.  The transformation between the variables $\sigma_i$ and $\sigma'_i$ is nonlocal: the operator $\sigma'_i$ flips a whole half-line of $\sigma$ spins, ending at $i$.  We will not derive this transformation or give its form explicitly here, but we will derive a similar transformation in \S2.3 below.

When $K$ is large, the action strongly favors all spins being parallel, and the system is ferromagnetic.  The duality then implies that for small $K$, the dual variables $\sigma_i'$ behave ferromagnetically; this is a disordered state in terms of the $\sigma_i$.  There must be a transition between these phases as $K$ is varied.  If there is only a single transition, then the duality implies that it must take place at the self-dual value $K =  K' = \frac12 \ln(1+\sqrt 2)$.  This is the case for the Ising model, but more general dual systems can have multiple transitions, which must occur in dual pairs if they are not at the self-dual point.

\subsection{Bosonization}

A second example of a surprising equivalence also arises in two spacetime directions.  A nice review of this equivalence is given by Coleman~\cite{Coleman:1974bu}.  One description is a massive Dirac fermion with a self-interaction,
\be
S_{\rm ferm} = \int d^2 x \left( i \bar \psi \gamma^\mu \partial_\mu \psi - m \bar \psi \psi - \frac{g}{2} \bar \psi \gamma^\mu \psi \,\bar \psi \gamma_\mu \psi \right)\,.
\ee
The other description is the sine-Gordon model, a scalar field with a cosine potential,
\be
S'_{\rm bos} =  \int d^2 x \left( - \frac12 \partial_\mu \phi \partial^\mu \phi + m \cos \beta \phi \right)\,.
\ee
In the bosonic description, $\beta$ plays the role of a coupling constant. One sees this by expanding the potential in powers of $\beta$:
\be
-m \cos \beta \phi = -1 + \frac{m}{2} \beta^2 \phi^2 - \frac{m}{24} \beta^4 \phi^4 \ldots \, .
\ee
Aside from the unimportant constant term, the leading term at small $\beta$ is a quadratic mass term, while interactions of the $\phi$'s are suppressed by additional powers of $\beta$.

Remarkably, these theories are equivalent, with the mapping of parameters
\be
\frac{\beta^2}{4\pi} = \frac{\pi}{\pi + g}
\ee
As $g$ goes to $\infty$, $\beta$ goes to zero.  As $\beta$ goes to $\infty$, $g$ goes to $-\pi$.  A negative $g$ means an attractive interaction, and $g = -\pi$ is the most attractive possible coupling, beyond which the theory is unstable.

This duality can be demonstrated by starting with the free massless theories, $m = \lambda = 0$.  One then finds by explicit calculation that with the correspondence
\be
\bar \psi \gamma^\mu \psi \leftrightarrow \frac{\beta}{2\pi} \epsilon^{\mu\nu} \partial_\nu \phi \,, \quad
\bar\psi \psi \leftrightarrow \cos\beta\phi \,, \label{dict}
\ee
the amplitudes in the two theories are equal.  One can then use the dictionary~(\ref{dict}) to show that deforming the fermionic theory toward nonzero $m$ and $g$ produces corresponding deformation of the bosonic theory.  (To do this all properly requires some attention to the proper definition of the renormalized operators in the quantum theory.)

It is not so surprising that one can make a boson out of two fermions.  Making a fermion out of bosons is more complicated.  The fermion is a {\it kink}, a configuration where the bosonic field $\phi$ takes different values in the two spatial limits $x^1 \to \pm \infty$.  Note that the multiple minima of the cosine potential, at $\phi = 2 \pi n /\beta$, make such a configuration stable.  As with the disorder operator in the Ising model, the basic quantum of one description is nonlocal in terms of the other.  More precisely, it is topological, being associated with nontrivial boundary conditions at infinity.

Both the Ising model and bosonization are realized in physical condensed matter systems.  For example, bosonization has been a valuable tool in understanding tunneling processes at the 1+1 dimensional edge of the quantum Hall system.  More generally, duality is often a powerful tool in condensed matter physics; some characteristic examples are~\cite{Shahar,Senthil}.

\subsection{Maxwell theory}

Our third example is Maxwell theory in four spacetime dimensions.  We begin with the sourceless theory, with action
\be
S_{\rm Maxwell} = - \frac{1}{4e^2} \int d^4x \,(\partial_\mu A_\nu -  \partial_\nu A_\mu)(\partial^\mu A^\nu -  \partial^\nu A^\mu)   \label{maxact}
\,.  
\ee
We use the conventional $e$ to denote the coupling, while $g$ will be used for the magnetic coupling later.
The path integral is over all vector potentials $A_\mu(x)$,\footnote{We might imagine that the fields are defined over some range of times $t_f > t > t_i$, with initial and final boundary conditions, so that the path integral defines a transition amplitude.  We will omit such details to focus on the central point.}
\be
\int {\cal D}A\, e^{i S_{\rm Maxwell}} \,. \label{zmax}
\ee 

Given an antisymmetric tensor $F_{\mu\nu}(x)$, the condition that it can be written as the curl of a vector potential is the Bianchi identity,\footnote{To be precise, this is true for a topologically trivial spacetime, as we assume here for simplicity.} 
\be
\partial_\mu \tilde F^{\mu\nu} = 0\,,\quad  \tilde F^{\mu\nu} = \frac{1}{2}\epsilon^{\mu\nu\sigma\rho} F_{\sigma\rho} \,,
\label{bianchi}
\ee
where $\epsilon^{\mu\nu\sigma\rho}$ is the fully antisymmetric Levi-Civita tensor. We can then replace the path integral over vector potentials $A_\mu(x)$ with a path integral over an antisymmetric tensor field $F_{\mu\nu}(x)$ subject to the constraint~(\ref{bianchi}) at each point,\be
\int {\cal D}A \ldots = \int {\cal D}F \prod_x \delta( \partial_\mu \tilde F^{\mu\nu}(x))\ldots \,. \label{atof}
\ee
There may be a Jacobian for this change of variables, but it is an uninteresting overall constant; similar constants are ignored in later equations as well.  Now write the functional delta-function in integral form, analogous to $\int_{-\infty}^\infty dx\, e^{i x y} = 2\pi \delta(y)$,
\be
\prod_x \delta( \partial_\mu \tilde F^{\mu\nu}(x)) = \int {\cal D} V \exp \left({ \frac{ i}{2\pi} \int d^4x\, V_\nu \partial_\mu \tilde F^{\mu\nu} }\right) \,. 
\label{deltint}
\ee
The factor of $1/2\pi$ is arbitrary; it can be absorbed into the normalization of $V_\mu$, but has been inserted to make later equations simpler.

Using the relations~(\ref{atof}) and~(\ref{deltint}), the path integral~(\ref{zmax}) becomes
\be
\int {\cal D}F\,{\cal D}V\, \exp \left\{-i \int d^4x \left( \frac{1}{4e^2}F_{\mu\nu} F^{\mu\nu} -  \frac{1}{4\pi} (\partial_\mu V_\nu - \partial_\nu V_\mu) \tilde F^{\mu\nu} \right) \right\} \,. \label{zmax2}
\ee
In the second term we have integrated by parts and made use of the antisymmetry of $\tilde F_{\mu\nu}$.  Now, integrating over $V_\mu$ just produces the functional delta function and takes us back to our starting point.  But integrating over $F_{\mu\nu}$ leads us to a new form.  This path integral is gaussian and can be carried out using the functional version of the usual identity
\be
\int dx\, e^{- i a x^2/2 +  i b x} = \sqrt{\frac{2\pi }{ia}} e^{ i b^2 /2 a} \,. \label{gauss}
\ee
The result, ignoring the normalization constant as before, is\footnote{Some useful identities are $\tilde S_{\mu\nu} \tilde T^{\mu\nu} = - S_{\mu\nu} T^{\mu\nu}$ for any antisymmetric tensors $S$ and $T$, and $\tilde{\tilde {T}\,}\!_{\mu\nu} = - T_{\mu\nu}$ for any antisymmetric tensor $T$. \label{tildefoot}}
\be
\int {\cal D}V\, \exp \left\{-{i \frac{ e^2}{16\pi^2}}  \int d^4x\,  (\partial_\mu V_\nu - \partial_\nu V_\mu)(\partial^\mu V^\nu - \partial^\nu V^\mu)  \right\} \,.
 \label{zmax3}
\ee
Comparing the forms~(\ref{zmax}) and~(\ref{zmax3}), we have turned a path integral over a vector field $A_\mu(x)$ into a path integral of similar form but over a new vector field $V_\mu(x)$, which entered originally as an auxiliary field in the integral representation of the functional delta function. 

To see the relation between these path integrals, consider the equation of motion from varying $F_{\mu\nu}$ in the action~(\ref{zmax2}),
\be
\tilde F_{\mu\nu} = - G_{\mu\nu}\,,\quad G_{\mu\nu} = \frac{e^2}{2\pi} (\partial_\mu V_\nu - \partial_\nu V_\mu) \,. \label{av}
\ee
That is, the electric part of $F = \partial_\mu A_\nu - \partial_\nu A_\mu$ is the magnetic part of $\partial_\mu V_\nu - \partial_\nu V_\mu$, and vice versa.  This shows that the electric-magnetic duality of free electromagnetism, $\vec{ E} \to \vec { B},\ \vec  { B} \to - \vec  { E}$ (the factor of $e^2/2\pi $ depends on conventions, and can be removed by rescaling fields).  

Note that the integral~(\ref{gauss}) is the Fourier transformation of a gaussian, illustrating the comment in \S1 that a duality is something like a Fourier transform.  Also, the relation~(\ref{av}) between $A_\mu$ and $V_\mu$ is nonlocal: it is local in terms of the curls of these potentials, but one must integrate over spacetime to express it in terms of the potentials themselves.

We can see already a very important lesson.  The path integral~(\ref{zmax}) has a gauge invariance
\be
A_\mu(x) \to A_\mu(x) + \partial_\mu \lambda(x)\,.
\ee
The path integral~(\ref{zmax3}) has a gauge invariance
\be
V_\mu(x) \to  V_\mu(x) + \partial_\mu \chi(x)\,.
\ee
Nothing transforms under the $\chi$ gauge invariance in the original theory~(\ref{maxact}, \ref{zmax}): this invariance is {\it emergent}.  
In the introduction we have noted that dualities call into question our notion of what is fundamental.  We see that this includes even gauge invariance, which we might have thought to be one of the fundamental principles.  This example might seem rather trivial, but the phenomenon of emergent gauge theories has proven to be much more general, ranging from condensed matter systems to string theory.

In perturbative gauge theories one seems find larger and larger gauge symmetries as one goes to higher energies: from the $U(1)$ of electromagnetism to the $SU(3) \times SU(2) \times U(1)$ of the Standard Model to the $SU(5)$ or larger of grand unified theories.  One might have supposed that the goal was to identify the ultimate gauge invariance from which all else descends.  We now recognize that in a variety of contexts gauge invariance can emerge from nothing.
This is consistent with the insight that gauge invariances are not symmetries, but rather redundancies, the same physical configuration being described by more than one field configuration.  Dualities relate only physical observables, and so are blind to gauge invariances.  Global symmetries, which act nontrivially on physical states, are the same in both dual descriptions.

In going from the original path integral (\ref{zmax}) to the dual~(\ref{zmax3}), $e$ has been replaced by 
\be
e' = \frac{2\pi}{e}\,.
\ee
This looks like the weak/strong duality that was discussed in \S1, but it is a fake.  This theory is free, with a gaussian action in both forms, and the `couplings' can be absorbed into rescalings of the fields $A_\mu$ and $V_\mu$.  We have written things the way we have in order to illustrate a more general principle, and we will note in \S2.4 some examples where analogous transformations produce a true weak/strong duality.

As a final exercise, consider adding a $\theta$-term 
\be
 \frac{\theta}{16\pi^2}  F_{\mu\nu} \tilde F^{\mu\nu} \label{thetaterm}
\ee
to the action~(\ref{zmax2}).  With a little algebra you can show that the final action in terms of $V$ can be written in terms of couplings $e'$ and $\theta'$.  The transformation is simple when written in terms of 
\be
\tau = \frac{\theta}{2\pi} +i\frac{2\pi}{e^2} \,. \label{tau}
\ee
It is simply 
\be
\tau' = -\frac1\tau\,.
\ee

\subsection{Generalizations}

The Maxwell theory can be generalized to $p$-form potentials, with $p$ antisymmetric indices, and to any number of spacetime dimensions $d$.  The exact same steps --- treat the field strength as a independent field, introduce the Bianchi constraint in integral form, and integrate out the field strength ---
again produce a dual theory, with a $p' = d-p - 2$ index potential (note that the fully antisymmetric $\epsilon$-tensor has $d$ indices).  We will refer to this general set of transformations as Hodge dualities.  The Maxwell case is $d=4$, $p = p' = 1$.  A variety of higher dimensional forms are present in string theory and supergravity.  A rather useful case is simply $d=2$, $p = p' =0$.  A $0$-form potential is just a scalar field, and its `field strength' is just its gradient.  This gives two representations of the two-dimensional massless scalar,
\be
\partial_\mu \phi = \epsilon_{\mu\nu} \partial^\nu \phi' \,. \label{ttrans}
\ee
The divergence of the right side is automatically zero, so we get a massless field equation for $\phi$.

The Ising model can be thought of as a scalar field theory in which the field lives on a discrete spacetime and is limited to discrete values $\pm 1$.
The exact same steps can be applied in the discrete case to produce a dual description, and this is actually the derivation of the Ising dual~(\ref{isingdual}).  In contrast to the Maxwell case, this is a true weak/strong duality: the discrete values of the variables don't allow for a rescaling.

The Ising model has $p$-form generalizations, where the variables live on $p$-dimensional cells of the lattice.  Again, the same steps produce a dual theory~\cite{Savit:1979ny}.  For example, the three-dimensional Ising model is dual to a three-dimensional $\mathbb Z_2$ gauge theory, where the potential lives on links and takes values $\pm 1$.  Further generalizations, again having duals, allow the potentials to take values in $\mathbb Z_N$, or the integers.  These systems have various phases, and the phase diagrams are duality-symmetric.

Consider now coupling the Maxwell theory to electrically and magnetically charged fields at the same time.  As shown by Dirac~\cite{Dirac:1931kp} , the quantum theory is  consistent only if the product of the electric charge $e$ and magnetic charge $g$ is a multiple of $2\pi$.  In particular, let us suppose that the product takes the minimal value, so that $g = 2\pi/e$.   Now apply the Maxwell duality above, which takes $e$ to $e' = 2\pi/e$ and switches electric and magnetic charges.  The electric charges $e$ become magnetic charges $e = 2\pi/e'$, while the magnetic charges $2\pi/e$ become electric charges $2\pi /e = e'$.  What this means is that the theories with couplings $e$ and $e'$ are {\it the same} under the Maxwell change of variables: we have found a self-duality.

This theory is somewhat unsatisfactory (and also somewhat obscure; we will meet a better one in the next section, in which the magnetic charges are solitons rather than independent fields).  It has no weakly coupled limit: the theory has both electric and magnetic charges, and in either description one of these will be strongly coupled.  Thus, we have found a `strong-strong' duality, which is not so useful.  Also, following the discussion in the introduction, this means that the theory has no classical limit --- whichever field is strongly coupled equivalently has large quantum fluctuations.    Related to this, it has no local and covariant Lagrangian: the Dirac quantization condition implies that the definition of the charged field depends nonlocally on the configuration of magnetic charges, and vice versa.\footnote{On a personal note, I have a long history with this subject.  The first warmup problem given to me as a graduate student by my advisor Stanley Mandelstam was to find a Lagrangian for this system.  I will explain his interest in this subject below.}
 A nonlocal Lagrangian is given in Ref.~\cite{Brandt:1978wc}.  This theory, with some additional supersymmetry, can be obtained as the low energy limit of a strongly coupled theory that does have a local covariant Lagrangian~\cite{Argyres:1995jj} .

\subsection{Montonen-Olive duality}

If one is keeping score to this point, we have described weak/strong dualities in two spacetime dimensions and for various discrete systems in higher dimensions, and a strong/strong duality in $d=4$, but no weak/strong duality for a continuum QFT in $d=4$.  Non-Abelian gauge theories are a place to look for this, because they have weakly coupled limits, and they can have magnetic monopoles.  These monopoles, found by 't Hooft and Polyakov~\cite{'tHooft:1974qc,Polyakov:1974ek}, arise as solitons, classical solutions to the field equations (like the sine-Gordon kink).

The non-Abelian electrically and magnetically charged particles (which we will refer to as charges and monopoles, respectively) seem to have very different origins: the former arise from the quantization of the fields, while the latter are particle-like solitons even in the classical limit.  At weak coupling, the electrically charged objects are light, weakly interacting, and essentially pointlike, while the magnetic monopoles are heavy (the soliton mass contains a factor of $1/g^2$ as compared to the electrically charged quanta), strongly interacting, and have finite radii that is set by the scale $v$ of gauge symmetry breaking.  Consider now what happens as the coupling is increased, as indicated by the orders of magnitude in the table.
\begin{table}[h]
\begin{center}
\begin{tabular}{rcc}
& electric & magnetic \\
& charges & monopoles \\ \hline
pair potential & $g^2$& $1/g^2$ \\
mass & $g v$ &  $v/g$\\
size$\,\times\,$mass ($g \ll 1$) & $\ll 1$ & $1/g$
\end{tabular}
\end{center}
\end{table}
The electrically charged objects interact more strongly, while the magnetically charged objects interact more weakly.  The ratio of the monopole to the charge mass goes down, and if the masses can be reliably extrapolated then for $g > O(1)$ the monopoles are actually lighter.  The electric quanta become less pointlike, as they split more frequently into virtual quanta.  The ratio of the monopole size to its Compton wavelength goes as $1/g$ and so becomes smaller.  If this last can be reliably extrapolated to large values $g$, then the monopoles end up much smaller than their Compton wavelengths.  In this case one would think that they can be quantized as fundamental fields.

In 1977, Montonen and Olive conjectured on this basis that non-Abelian gauge theories having magnetic monopole solutions would be invariant under a weak/strong, electric/magnetic duality; the simplest example would be the Georgi-Glashow $O(3)$ weak interaction model.  In the following year, Witten and Olive introduced supersymmetry into the story~\cite{Witten:1978mh}, showing that in a supersymmetric extension of this model the calculation of the masses is exact, so that at large $g$ one can reliably say that the monopoles are lighter than the charges.  To be precise, one needs ${\cal N}=2$ supersymmetry for this to hold.\footnote{In four spacetime dimensions, the smallest supersymmetry algebra has one Majorana or Weyl supercharge, meaning four real or equivalently two complex supersymmetry charges.  Some systems are invariant under extended supersymmetry, with ${\cal N}$ copies of this basic algebra.  Larger algebras require particles of higher spin, so that  ${\cal N}= 4$ is the maximum for a QFT without gravity, and ${\cal N=8}$ is the maximum for a gravitational theory.}   Osborn then showed that for ${\cal N}= 4$ supersymmetry (but not less), the spins of the charges and monopoles match as well.

In retrospect, this is strong circumstantial evidence.  In particular that fact that the monopoles are much lighter than the charges at strong coupling is striking.  However, this duality could not be demonstrated by the kind of Lagrangian manipulations employed above (in particular, the non-Abelian action and Bianchi identity contain $A_\mu$ and not just $F_{\mu\nu}$, so the trick of treating $F_{\mu\nu}$ as the independent field goes nowhere fast).

A Lagrangian derivation is still lacking, but in the 1990's the circumstantial evidence for such dualities expanded rapidly.  Sen showed that also the dyon spectrum (e.g.\ magnetic charge 2, electric charge 1) is as predicted by duality~\cite{Sen:1994yi}.  To understand this, we should note that there is more to the duality group than the weak/strong interchange.  Referring back to the $\theta$ term~(\ref{thetaterm}), the quantization of instanton charge implies an invariance $\theta \to \theta + 1$.  For the combined parameter $\tau$ (\ref{tau}), we then have the two invariances $\tau \to - 1/\tau$ and $\tau \to \tau+1$.  Together these generate the infinite discrete group~\cite{Cardy:1981qy}
\be
\tau \to \frac{a \tau + b}{c \tau + d} \,,\quad \left[ \begin{array}{c} \vec E \\ \vec B \end{array} \right] \to  \left[ \begin{array}{cc} a & b \\ c & d \end{array} \right] \left[ \begin{array}{c} \vec E \\ \vec B \end{array} \right] \,,
\ee
where $a,b,c,d$ are integers such that $ad-bc = 1$.  This symmetry, known as $SL(2,\mathbb Z)$, relates the spectrum of dyons to that of the electric charges.  Another check was the demonstration by Vafa and Witten that certain twisted versions of the path integral are dual~\cite{Vafa:1994tf}.

Beyond this accumulation of evidence for the ${\cal N} = 4$ theory, this subject exploded when Seiberg began to solve the strongly coupled dynamics of ${\cal N}=1 $ supersymmetric theories~\cite{Seiberg:1994bz}, and Seiberg and Witten did the same for ${\cal N}=2$ theories~\cite{Seiberg:1994rs}.    These provided many cross-checks on the reliability of these circumstantial arguments, and they showed that ${\cal N} = 4$ is not an isolated example but part of an elaborate and massively self-consistent web.
Not long after, this web was extended to string theory, as we will describe in \S3.  Even for a skeptic, accustomed to path integral arguments, it became impossible to resist the weight of evidence.  

Finally, we emphasize that as in the simple Maxwell duality, the gauge invariances in the two dual descriptions are unrelated.  From the point of view of either theory, the other gauge symmetry is emergent.   The dual theories may even have different Lie algebras~\cite{Goddard:1976qe}.

\subsection{Remarks}

The absence of a derivation remains a puzzle.  The path integrals on both sides can be given rather precise definitions, so we have a sharp mathematical statement (actually many statements, the matching of all correlators and other observables) and no proof.  It should be noted that in quantum field theory proof usually lags far behind what we can understand physically, but one feels that this is a gap that must eventually be closed.  
It may be that the Wilsonian path integral construction of field theory is not the correct starting point~\cite{cyberg}. To this day, it remains the only universal tool for proceeding from a Lagrangian or a Hamiltonian to the full quantum theory, but it seems clumsy in theories of high symmetry.

A possible alternative would be to find a construction that applied only with a large amount of symmetry, and then reach theories of less symmetry  by some process of deformation.  In this context, we note an argument for $S$-duality of ${\cal N}=4$ Yang-Mills theory, beginning with the $d=6$ $(2,0)$ theory~\cite{Witten:1995zh}.  When two of the six dimensions are made periodic, the low energy physics is the $d=4$, ${\cal N}=4$ theory.  Moreover,  the coupling $g$ is the ratio of the periods of the two dimensions.  But there is no difference between these two directions, so by switching them we invert the coupling!  But what is this $(2,0)$ theory?  Well, it does not have a Lagrangian, or a classical limit, or any complete definition.  It is inferred to exist from the low energy limit of the physics on certain stringy 5-branes.  So clearly much is lacking, but this may point the way to the future, if we could construct this theory.  Further, a host of ${\cal N}=2$ dualities follow by compactifying it on other two-dimensional manifolds~\cite{Gaiotto:2009we}. 
 
What is the role of supersymmetry in $S$-duality?  Supersymmetry allows the calculation of certain quantities at strong coupling, and so makes it possible to check duality conjectures where there are  no other methods.  But in many cases, supersymmetry plays a more essential role: it allows the strongly coupled theory to exist at all.  To see the issue, consider adding to the QED vacuum an electron-positron pair in some region of size $r$ smaller than the Compton wavelength of the particles.  This costs a kinetic energy of order $1/r$, times two.  However, there is a negative potential energy of order $-e^2/r$.  For $e^2$ sufficiently greater than 1, adding the pair lowers the energy.  The would-be vacuum is then unstable at all scales $r$, and there may be no final state at all.  But in supersymmetric theories, there is a natural stability built in.  Essentially the Hamiltonian is the sum of squares of  supersymmetry charges, $H = \sum_i {\cal Q}_i^2$, and so the energy is nonnegative and bounded below.

In general, as the amount of supersymmetry is reduced, the dynamics becomes richer, and dualities act in more complicated ways.  Duality symmetries of some form exist in nonsupersymmetric theories, but generally they apply only in the low energy limit.  As a familiar such example, the low energy limit of QCD is the theory of pion physics, which is also described by a sigma model in terms of scalar fields.

Finally, we note another interesting application of duality.  It was observed by 't Hooft~\cite{'tHooft:1977hy} and Mandelstam \cite{Mandelstam:1978ed} that confinement of quarks (another universally believed but unproven property in quantum field theory) would follow if the QCD vacuum were the electric/magnetic dual of a superconductor.  A superconductor excludes magnetic fields via the Meissner effect, so a dual superconductor would exclude (color) electric fields and  force color-electric charges into neutral bound states.  QCD itself does not have a precise dual, but this mechanism can be seen in various supersymmetric theories that are closely related to QCD~\cite{Seiberg:1994rs}.

\sect{String/string dualities}
\setcounter{equation}{0}

In string theory we need to distinguish two kinds of quantum effects: the wiggles on a single string, and the splitting of the string into two strings.  These are controlled by different parameters.  The importance of the wiggles depends on $l_{\rm s}/ l$, where $l_{\rm s}$ is the characteristic length scale of string theory, and $l$ is the characteristic length of the system being studied.  The splitting depends on the string coupling $g_{\rm s}$.  Each quantum effect in turn is associated with a kind of duality, known as $T$- and $S$-duality respectively.  The $T$-duality is simpler, since the wiggles of the string are described by a quantum field theory, though we will see that it teaches us rich lessons.  The $S$-duality gets into the full and puzzling nature of quantized string theory.

\subsection{$T$-duality}

String theory originally took hold as a solution to the short distance problem of quantum gravity.  If we try to build a quantum theory of gravity by the standard procedure of feeding the classical Lagrangian into the path integral, the result is nonrenormalizable: the divergences become worse with each order of perturbation theory.  Such problems had been encountered before, and had pointed the way to new physics.  The Fermi theory of the weak interaction was highly successful in accounting for observations, but it too was nonrenormalizable, pointing to a breakdown of the theory at short distance.  In order to solve this problem, it was necessary to resolve the pointlike interaction of the Fermi theory into the exchange of a $W$ boson.  Indeed, this clue, together with some imagination, was enough to lead to the Weinberg-Salam model.  This predicted the precise properties of the $W$ and $Z$ bosons, and the Higgs, well before there was any direct evidence that any of these existed.  

In the case of gravity, the new physics has to be more than just a few new particles.  In string theory, the basic point-like quanta of QFT are expanded into loops and strands.  This smooths the short distance behavior of the amplitudes.  But going from points to strings took us out of the familiar framework of QFT, where we have had many decades to  learn how things work.  For string theory, we had rules of calculation, but not the intuition for all that they imply.  In this circumstance, it is useful to consider a variety of thought experiments.
A particular thought experiment that has been enormously fruitful is imagining what happens if we put the strings in some compact space, and then contract the space.  

For comparison, let us do this first for a quantum field.   We will take the mathematically simplest kind of box, where one or more dimensions are  periodic with period $2\pi L$.  For a field to fit into such a box it must be of the form $e^{i n/L}$ for integer $n$.  This corresponds to a momentum $n/L$, and therefore an energy 
\be
E^2 = p_\bot^2 + \frac{n^2}{L^2} + M^2\,, \label{parte}
\ee
where $p_\bot$ is the momentum in the noncompact directions and  $M$ is the rest mass.  Now, as we take $L \to 0$, this energy diverges.  The only states of finite energy are those with $n=0$, those that don't depend on the compact dimension at all.  If we start with $d$ spacetime dimensions, and compactify $k$ of them in this way, then only the $d-k$ noncompact momenta can be zero: the fields move only along the noncompact dimensions.  In effect, the small dimensions disappear,  at least to probes with energy less than $1/L$.

For a closed string (a loop), the energy is
\be
E^2 = p_\bot^2 + \frac{n^2}{L^2} +  \frac{w^2 L^2}{\alpha'^2} + \frac{\pmb N}{\alpha'} \,. \label{stringe}
\ee
The first and second terms are the same as in the QFT energy~(\ref{parte}).  The final terms are also the same: the mass-squared of a string is proportional to its excitation level $\pmb N$ (which may include a zero-point constant), and inversely proportional to $\alpha'$.  The third term in the string energy~(\ref{stringe}) is special to string theory.  It arises because a closed string can wind around the periodic dimension.  The integer $w$ counts the number of windings.
This energy comes from the tension of the stretched string, and so is proportional to the distance $2\pi L$ that the string must stretch.
 
Now consider what happens as $L \to 0$.  The states of nonzero $n$ again go to high energy.  But now we have a large set of very low energy states at nonzero $w$.  This is the opposite of what happens in the $L \to \infty$ limit, where the $n \neq 0$ states go over to a continuum while the $w\neq 0$ states go to high energy.  In fact, the energy is symmetric is completely symmetric between large $L$ and small $L$: if we interchange
\be
L \to \frac{\alpha'}{L} \,,\quad  (n ,w) \to (w, n)
\ee
it is unchanged~\cite{Kikkawa:1984cp,Sakai:1985cs}.  Evidently there is no difference between large $L$ and small $L$!  This holds not just for the spectrum~(\ref{stringe}), but for the interactions of the string as well~\cite{Nair:1986zn}.  In the effective two-dimensional QFT of the string world-sheet, $T$-duality is just the  $d=2$, $p = p' =0$ Hodge duality~(\ref{ttrans}).  The coordinates of the string in spacetime are scalar fields in the world-sheet QFT, and the coordinates in the dual theory are related as in Eq.~(\ref{ttrans}).

This comparison of strings~(\ref{stringe}) with point particles~(\ref{parte}) reveals that strings see spacetime differently.  In particular we see signs of a minimum length at the self-dual value $L = \sqrt{\alpha'}$.  This is also an example of emergent spacetime, to go along with the emergent gauge symmetry noted earlier: the effective large spacetime that appears as $L \to 0$ is not the original one in which the string was embedded, but emerges from the quantum dynamics of the string.

There is a natural question here: if large $L$ and small $L$ are equivalent (e.g. a light-year and $10^{-77}$ cm), why should we prefer the large-$L$ picture?\footnote{I thank Eliezer Rabinovici for raising this question and sharing his thoughts on it.}  The point is that there is a locality structure, things far apart in the large-$L$ picture cannot interact directly with each other.  This important property is not manifest in the small-$L$ picture.

Beyond the simple $T$-duality of periodic flat dimensions, stringy geometry extends to curved spaces in interesting ways.  Mirror symmetry is an extension of $T$-duality relating strings moving on different Calabi-Yau manifolds.  It implies surprising connections between different mathematical objects, and has led to extensive interaction between mathematics and string theory~\cite{Hori:2003ic}.  These methods have also led to controlled descriptions of transitions between spaces of different topology, going beyond familiar notions of geometry.  Further developments are described in the reviews~\cite{Giveon:1994fu,Quevedo:1997jb}.

It is very interesting to extend this thought experiment to open strings~\cite{Dai:1989ua,Horava:1989ga}.  Open strings do not have a winding number: their ends are free, so they can simply unwind.  Correspondingly, the mass formula 
\be
E^2 = p_\bot^2 + \frac{n^2}{L^2}  + \frac{\pmb N}{4\alpha'} \,. \label{ostringe}
\ee
has no winding number term.  As $L \to 0$, no new light states emerge, and the small dimension disappears just as in \mbox{QFT}.  But this is a puzzle.  Theories with open strings always have closed strings as well.  So if we start with open plus closed strings in $d$ spacetime dimensions, and take $k$ dimensions to be very small, the closed strings still feel a $d$-dimensional space, but the open strings move in $d-k$ dimensions.  The physical interpretation of this calculation, as can be can be confirmed by further thought experiments, is that in the dual picture we do not have empty space, but rather a membrane with $d-k-1$ spatial dimensions, and the endpoints of the open strings are stuck to this manifold. These  were termed D-branes (D for Dirichlet, referring to the boundary condition for the open string endpoints).

The surprising lesson is that string theory has additional extended objects of various dimensions.  One might think of them as being solitons, 
built out of strings in some way.  More precisely, they are open string solitons.  Around the same time, other classes of extended objects built out of closed strings were found --- black branes~\cite{Horowitz:1991cd} and NS5-branes~\cite{Callan:1991dj}.  The precise role of all of these took a few years to become clear, and we will return to it.

The search for consistent theories of relativistic one-dimensional objects eventually led to five distinct string theories: types I, IIA, and IIB,  and heterotic string theories with gauge groups $SO(32)$ and $E_8 \times E_8$.  All of these are supersymmetric, and in all cases consistent quantization requires that the string move in ten dimensions.
These actually involve just two kinds of string.  In types I, IIA, and IIB, supersymmetric degrees of freedom move along the string in both directions.  In the two heterotic theories, supersymmetric degrees of freedom move in one direction, and gauge degrees of freedom in the other.  The differences between the theories in each class are the boundary or periodicity conditions on a piece of string of finite length.

In fact, $T$-dualities connect the theories within each class.  A $T$-duality on type IIA produces IIB, and vice versa~\cite{Dine:1989vu,Dai:1989ua}: the interchange of winding with momentum has this effect on the periodicity conditions.  A $T$-duality on the type I theory produces a space with a D-brane immersed in a IIA or IIB background, as described above.\footnote{The type I theory itself can be thought of as a type II theory with D9-branes.  That is, they fill the nine space dimensions, so are not perceived as dynamical objects.  The field equations require that there be 16 D9-branes, as well as a related object, the orientifold 9-plane, which together lead to a gauge group $SO(32)$.}  For either heterotic theory, if one takes the $L\to 0$ limit in combination with turning on a background gauge potential, one can get to the other~\cite{Narain:1985jj}.

In all, studying strings in small spaces has been a remarkably productive thought experiment, leading to stringy geometry, D-branes, and connections between seemingly distinct string theories.

\subsection{$S$-dualities}

We now consider dualities with respect to the string coupling $g_{\rm s}$.  As with the field theory $S$-dualities, an effective strategy is to look at quantities that can be calculated at strong coupling using supersymmetry, and comparing with the weak-coupling values.  These quantities include the low energy effective theory (the spectrum and interactions of the massless fields), and the spectrum of BPS states.  BPS states are massive states that are invariant under some subgroup of the supersymmetry algebra.  To see the basic idea~\cite{Witten:1978mh}, a typical supercharge $Q$ has the property
\be
Q^2 = H - G \,,  \label{centralcharge}
\ee
where $H$ is the Hamiltonian and $G$ is some other conserved quantity; $G$ may depend on the couplings, but this dependence is determined by supersymmetry).  Then for any state,
\be
\langle\psi | H |\psi \rangle - \langle\psi | G |\psi \rangle = \langle\psi | Q^2 |\psi \rangle = \| Q  |\psi \rangle \|^2 \geq 0 \,.
\ee
Equality holds if and only if $Q  |\psi \rangle =0$ for one or more supercharges, in which case $ |\psi \rangle$ is termed a BPS state.  The energy of such a state is thus determined by its charge, for any value of the coupling.

Evidence for string-string $S$ dualities appeared as early as Refs.~\cite{Cremmer:1978km,Duff:1987bx,Font:1990gx,Duff:1994zt}, but the full and intricate pattern only emerged with Refs.~\cite{Hull:1994ys,Townsend:1995kk,Witten:1995ex,Horava:1995qa}.  Consider first IIB string theory.  This has odd-dimensional D-branes, so there are two one-dimensional BPS objects, the fundamental string whose quantization defines the theory, and the D1-brane or D-string.\footnote{The second object was originally described as a `black brane.'  The relation with the D-brane picture will be explained in \S4.}
Their tension is in the ratio 
\be
\tau_{\rm F1}/\tau_{\rm D1} = g_{\rm s} \,.
\ee
This is invariant under $g_{\rm s}  \to 1/g_{\rm s}$ with interchange of the F1 and D1, suggesting self-duality of  IIB string theory.  This is also consistent with various other evidence, such as the self-duality of the low energy effective action.  It is interesting to look at the effect of the duality on D3-branes.  A characteristic property of D3-branes is that their massless degrees of freedom are described by an ${\cal N}=4$ gauge theory, whose gauge group is $U(N)$ for $N$ coincident D3-branes~\cite{Witten:1995im}.  The IIB duality takes D3-branes into themselves, but it reverses the electric and magnetic fields on them.  Thus, IIB self-duality implies Montonen-Olive duality.  The reverse is not necessarily true, but this is one simple example of the web of interconnections among all the dualities.

The IIA string has even-dimensional D-branes, so no D-string.  One clue as to its dual was the early observation~\cite{Cremmer:1978km} that  Kaluza-Klein (KK) compactification of $d=11$ supergravity gives the massless sector of the IIA string theory.  A second clue~\cite{Duff:1987bx} was existence of 2-brane solutions in $d=11$ supergravity.  In KK compactification, a 2-brane with one direction wrapped on the KK circle behaves exactly like a IIA string.  The IIA D-branes also play essential roles~\cite{Hull:1994ys,Townsend:1995kk,Witten:1995ex}.  The D0-brane, a massive particle, corresponds to the charged states in the KK compactification, while the D2-brane is the $d=11$ 2-brane with both directions orthogonal to the IIA circle.  The higher-dimensional D-branes each have their $d=11$ interpretation in turn.  

These arguments suggest the existence of an eleven-dimensional theory of quantum gravity, whose low energy limit is $d=11$ supergravity.    The maximum dimension for a weakly coupled, Lorentz-invariant, theory of strings is $d=10$.  The perturbative expansion of the IIA amplitudes in powers of $g_{\rm s}$ is an expansion around the zero-radius limit of the KK compactified $d=11$ theory, so the eleventh dimension is not visible at weak coupling.  The $d=11$ theory is not a perturbative string theory, but it is one limit of a theory of quantum gravity, which includes string theories as other limits.  The full form of this quantum theory is not yet known; it has been given the provisional name of M theory~\cite{Witten:1998uk}.  (Sometimes M theory is used specifically for the $d=11$ theory, and the branes in this theory are termed M-branes.)

The type I string does have a D1-brane~\cite{Polchinski:1995df}, which is heterotic: supersymmetric degrees of freedom move along it one direction, and gauge degrees of freedom move in the other direction.  These are the same properties as for the fundamental string of the heterotic theory.  Moreover the gauge group of the type I theory is $SO(32)$, the same as for one of the two heterotic theories, and so we identify the type I and $SO(32)$ heterotic theories as weak/strong duals of one another.

The remaining string theory, the $E_8 \times E_8$ heterotic theory, took the longest to sort out~\cite{Horava:1995qa}, but by using its $T$-duality to the $SO(32)$ theory, and following a further chain  of dualities, one can show that it is also dual to a compactification of the $d=11$ theory.  Normal KK compactification on a circle gives the IIA theory, but it turns out that compactification on a line segment, with a boundary at either end, is also consistent and gives the $E_8 \times E_8$ heterotic theory.

In all, this is a rather nontrivial pattern: two string theories are $S$-dual to each other, one is $S$-dual to itself, and two are $S$-dual to compactifications of a $d=11$ theory.  Moreover, by combining this with connections via $T$-duality described earlier, one sees that one can get from any string theory to any other by a series of $T$- and $S$-dualities.  The situation is shown schematically in the figure.
\begin{figure}[!ht]
\begin{center}
\vspace {-5pt}
\includegraphics[width=6in]{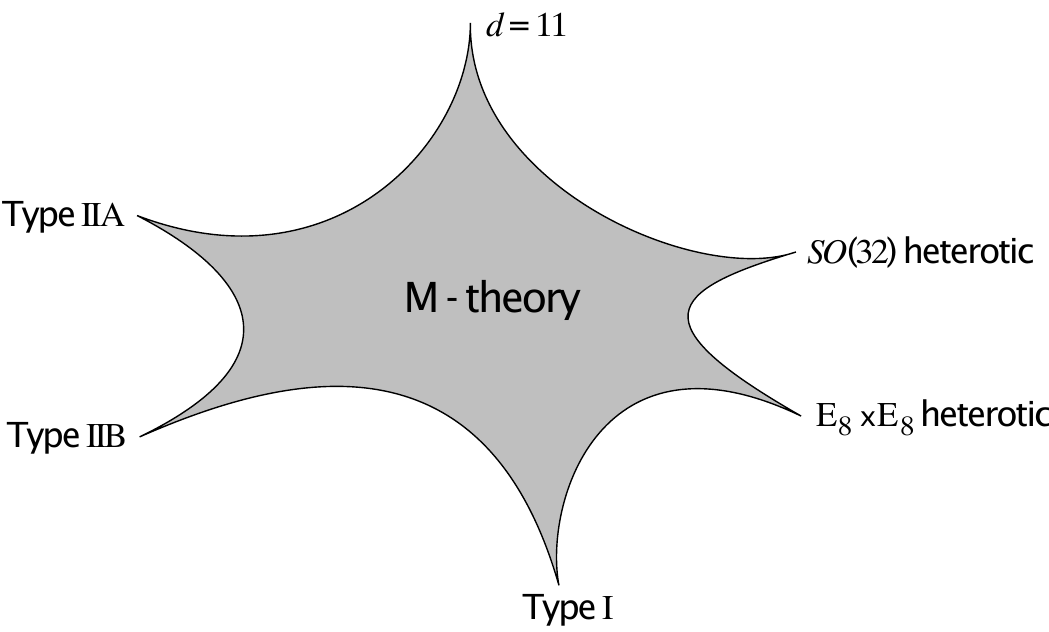}
\end{center}
\vspace {-10pt}
\caption{Schematic duality diagram.  A two dimensional subspace of supersymmetric string vacua, with the limits corresponding to the five string theories and $d=11$ supergravity labeled.
}
\label{fig:radii}
\end{figure}
This depicts a two-dimensional slice through the space parameterized by the string coupling and the radii of compact dimensions, and emphasizes the nature of string theory and the $d=11$ theory as limits of some single quantum theory.  Since string theory contains gravity, one might have thought that the strongly coupled limit would involve strong gravity, with the metric fluctuating wildly.  The dualities mean that if we try to reach such an extreme, we find instead a dual picture in which things become classical again.

There is an important point to emphasize here.  We have been referring to the different points within the gray region as different theories.  However, they are distinguished, not by the values of fixed constants, but by the values of background fields.  This is clear for the radii of dimensions, which arise from the metric, but it is also true for the string coupling, which is determined by the value of the dilaton field.  Thus, the figure depicts a single quantum theory with many backgrounds (vacua), and in certain limits the physics in the given background is well-described by one or another weakly coupled theory.

In QFT a theory is first characterized by the choice of fields and symmetries, with an infinite number of options.  In string theory, we only have five choices for the kind of string, and duality shows them all to be equivalent.  But secondly, in QFT for each set of fields and symmetries there may be a continuous infinity of possible coupling constants.  In string theory, this second infinity comes from the Hilbert space.  It is a property of the state rather than the theory.  This is not a consequence of duality per se. Rather it reflects the Einstein-Kaluza-Klein principle that physics emerges from geometry.

The space of vacua shown in the figure is a tiny piece of the landscape of string theory.  The landscape follows from a rather simple train of thought.  
General relativity describes gravity as the curvature of space and time.  In a unified theory, we would expect this to be true for the other forces as well. But for this we need `more spacetime,' essentially more dimensions: this is Kaluza-Klein theory.  String theory requires extra dimensions for an additional reason, the consistent quantization of the string.  In string theory the Kaluza-Klein mechanism can be transmuted into many other forms by dualities, but the central idea remains, that the physics that we see reflects the spacetime geometry.  The equations of general relativity in higher dimensions have multiple solutions, corresponding to multiple vacua in the quantum theory.  The vacua in the figure are all degenerate, a consequence of the large amount of supersymmetry, but for nonsupersymmetric solutions there is a potential energy that leads to isolated solutions.  Rather than the two dimensions shown in the figure, topologically complex solutions are characterized by hundreds of background fields or more, and combinatorics can generate a vast number of solutions.

The physics that we see directly depends on the shape of the compact directions,  not just on the underlying equations of string theory.  The latter seem to be unique, but the former are far from unique.  The uniqueness of the equations is what one would hope for from a complete theory.  The nonuniqueness of the solutions is again in keeping with the general pattern in physics, that a few principles can define a theory, but these then give rise to a vast range of different phenomena.  The consequence here is that things that we had hoped to predict, basic constants and spectra of low energy physics, depend on which of the many vacua we are in.  But there is some evidence that this is the way things are. Weinberg argued in 1987 that such a multi-vacuum theory would explain the absence of a large cosmological constant and predict a small nonzero one~\cite{Weinberg:1987dv}, and after 27 years and the discovery of vacuum energy there is still no significant competing idea.

We should note one important difference between string theory and field theory.  In field theory, we have the Hamiltonian~(\ref{ham}), and we have the corresponding perturbative expansion of the amplitudes~(\ref{perts}).  In string theory we had for a long time only the analog of~(\ref{perts}), the sum over string world-sheets.  The dualities are deduced, not from a full understanding of the exact theory, but from parts of it that are determined by general principles plus supersymmetry.   In QFT, the path integral as defined by Wilson tells us what is the exact theory.  We might try to copy this in string theory, in the form of an infinite component QFT known as string field theory, but this has not been as fruitful as hoped and another approach seems needed.  The string-string dualities do give us some more global picture of the theory, and we will see in the next section that further dualities give a much fuller understanding.

\sect{QFT/string dualities}

Dualities between field theories, and dualities between string theories, are remarkable, taking QFT and string theory far beyond their perturbative descriptions.  A duality between a field theory and a string theory might seem to be impossible, on several grounds.  String theories require ten dimensions, whereas renormalizable field theories do not seem to exist in dimensions greater than four (though some non-Lagrangian theories exist up to dimension six).  String theories seem to contain many more degrees of freedom than QFT's, from the infinite number of internal states of the string.   And, string theories contain quantum gravity, with its many conceptual puzzles, while renormalizable QFT's do not.  Well, prepare to be amazed.

\subsection{Matrix theory}

The short distance problem of quantum gravity points to a minimum spacetime distance.  This is suggestive of some sort of uncertainty principle for position,
\be
[ X^i, X^j ] \neq 0 \,, \label{xxcom}
\ee
extending the Heisenberg relation~$[ p^i, x^j ] = -i\hbar$.  Here $i, j$ are spatial indices; we will get to Lorentz invariance below.  Now, if we try to copy the Heisenberg relation and put some constant on the right side of~(\ref{xxcom}), we have a problem with rotation invariance: the left-hand side has two antisymmetric spatial indices, so a constant of this form would not be rotationally invariant.  So we will try a different approach.  We will just assume that the $X^i$ are $N \times N$ matrices in some unspecified space, and figure out what this means later.  

We want to have ordinary commutative space at low energies, so we will include in the Hamiltonian a simple term that enforces this.  We take
\be
H = {\rm Tr}\left( \dot X^i \dot X^i +| [X^i, X^j] [X^i, X^j]|\right) \,. \label{mham}
\ee
For convenience, all coefficients have been set to 1.
The first term in $H$ is a kinetic energy; expanding out the trace, one gets the sums of the squares of the time derivatives of all of the matrix components (we use summation convention for the repeated spatial indices).  The second term vanishes only if all the $X^i$ matrices commute.  In this limit, we can diagonalize these matrices, and the diagonal elements just give us $N$ free particles.  However, at high energy the full noncommutative structure comes into play.    In addition we have to introduce supersymmetry, for the usual reason of stability, so we add in fermionic coordinates $\Theta$ in an appropriate way.   This adds one more term to $H$, a Yukawa interaction $\Theta \Theta X$.

It might seem that we have just made a toy model of noncommutative spacetime, but in fact we are done: the quantum mechanical model we have described is a dual description of M theory!  This Hamiltonian first made its appearance in attempts to quantize the M2-brane as a fundamental object, by analogy to the quantization of the string in string theory~\cite{de Wit:1988ig}.  The M2-brane, having one more dimension than the string, has more internal degrees of freedom and these give rise to more divergences; the finite value of $N$ corresponds to a regulated version of the M2-brane.  Of the eleven dimensions of M theory, nine come from the matrices $X^i$ with $1 \leq i \leq 9$; this is the maximum number allowed by the supersymmetry.  Time provides a tenth dimension, and the eleventh is hidden here in Fourier-transformed form: $N$ is the number of units of momentum in this direction.  The quantization of the membrane is in light-cone gauge, which treats the dimensions in this asymmetric way.\footnote{There is a covariant form of Matrix theory, based on ten matrices $X^\mu$, which is suppose to describe IIB string theory~\cite{Ishibashi:1996xs}.  However, its full interpretation is not clear.  One issue is that time, which is one of the ten matrices, must be Euclidean.}

The full significance of this Hamiltonian was identified by Banks, Fischler, Shenker, and Susskind~\cite{Banks:1996vh}: it describes not just the membrane, but the whole of M theory, at a particular location in the landscape of Fig.~1.  In particular, the Hamiltonian is precisely the Hamiltonian of D0-branes, which are interpreted as eleven-dimensional gravitons in the IIA-M $S$-duality.  The BFSS Matrix theory corresponds to M theory with one of its dimensions compactified in a null direction, and with $N$ units of momentum in this direction.  But even better, it allows one to describe the quantum theory of M theory in eleven uncompactified dimensions as well.  By taking $N \to \infty$, one is essentially taking the compact dimension to be large, in the rest frame of the matter, and in the limit one gets the noncompact theory.  This is the upper point in the duality diagram, but unlike the supergravity description, which captures only low energy physics, the Matrix theory description is complete.

This means that if we want the $S$-matrix in M theory, we need simply solve the quantum mechanics problem~(\ref{mham}) and take the limit $N \to \infty$.  Now, this is hard, but it is something that one could program a very large computer to do, and so it provides an algorithmic definition of the theory.  In fact, similar calculations are being done~\cite{Hanada:2013rga}.  They are near the limit of current technology, but show impressive agreement between matrix quantum mechanics and quantum gravity.

The simple quantum model that we have described thus provided something that was previously lacking, a nonperturbative construction of at least part of the landscape of vacua of string theory~\cite{Banks:1996vh}.  One challenge that remains here is that if one compactifies some of the dimensions (to get down to the four noncompact dimensions of our vacuum), Matrix theory becomes more complicated, and if more than three dimensions are compactified its form is not known.  Other challenges will be discussed below.

Regarding our earlier puzzle about the mismatch in the number of degrees of freedom, the key is that the full duality emerges in the $N \to \infty$ limit, and one can put as many degrees of freedom as one wants in an infinite dimensional matrix.  This seems like a truism, but the miracle is that the simple Hamiltonian~(\ref{mham}) is all that it takes to do the encoding.  To be precise, this is a string-QM duality, not a string-QFT duality.  But for uniformity with other examples that we will soon see, we should regard it as  a 0+1 dimensional field theory, the basic quantum variables $X$, $\Theta$ being functions only of $t$.

\subsection{Black hole quantum mechanics}

We have seen that one thought experiment, the string in a small box, was exceedingly productive.  For quantum gravity, black holes turn out to be another valuable laboratory, leading to the entropy puzzle and the information paradox.

\subsubsection{Black hole entropy}

By considering a process of feeding quantum bits into a black hole, Bekenstein argued that black holes have a well-defined information-carrying capacity.  By the uncertainty principle, it requires an energy $\hbar c/R$ to fit a quantum into a black hole of radius $R$.  Using the mass-radius relation for a black hole, $M = c^2R/2G$, one finds that a black hole of radius $R$ can contain of order $c^3 R^2/\hbar G$ bits of information.

This argument was reinforced by the discovery of Hawking radiation~\cite{Hawking:1974sw}.  Black holes radiate like a hot body with a temperature of order $\hbar c/R k_{\rm B}$.  By thermodynamic relations this translates into an entropy
\be
\frac{S}{k_{\rm B}} = \frac{\pi c^3 R^2}{\hbar G} = \frac{A}{4 l_{\rm P}^2} \,. \label{bhe}
\ee
In the last form, $A$ is the area of the black hole horizon, and $l_{\rm P}$ is the Planck length $1.6 \times 10^{-32}$ cm.  In statistical mechanics, the entropy is a count of the effective degrees of freedom of a system, and the number of possible microscopic states is the exponential of this.  Since $l_{\rm P}$ is very small, the number of states is very large, for macroscopic $A$.

This is a puzzle: what is the black hole entropy counting?  In general relativity, every black hole of given mass (and charge and angular momentum) looks exactly the same --- black holes have no hair.  But in quantum theory, they seem to have a microscopic structure.  Heuristically, temperature normally comes from the motion of atoms, so black holes should have some sort of `atomic' structure.

The first concrete description of the degrees of freedom of a black hole came from the work of Strominger and Vafa~\cite{Strominger:1996sh}.  They considered a thought experiment in which one starts with a black brane (like a black hole, but extended in some directions) and then reduces the strength of the gravitational interaction.  As we have noted, in string theory the coupling constant is the value of a field, the dilaton, and one can imagine reducing this in a gradual way, until the black brane is no longer black and we can see what is inside.

To do this in a controlled way, one wants to take a BPS black brane, so that one can keep track of its mass during the process.  For a particular class of black branes, what one finds inside is D-branes.  To understand this, recall that the supersymmetry algebra~(\ref{centralcharge}) contains a conserved charge $G$, and BPS objects carry this charge.  For D-branes, a key characteristic is that they carry Ramond-Ramond (RR) charge~\cite{Polchinski:1995mt}.  This kind of charge is not carried by fundamental string themselves, but for any D-brane, one can write down a `black' gravitational solution with the same number of extended directions and carrying the same charge~\cite{Horowitz:1991cd}.  This solution has a horizon and a singularity, and the Ramond-Ramond field lines emanate from the singularity.

Having two BPS objects with the same charge was a bit puzzling, until Ref.~\cite{Strominger:1996sh} observed that they were the weak and strong coupling descriptions of the same object.  Moreover, the weakly coupled D-brane picture allows an explicit count of the states of the system, and it agrees with that deduced by Bekenstein and Hawking.  

This subject, the count of black hole states, has continued to give insight into the structure of quantum gravity.  It points to a connection between spacetime geometry (the black hole area) and information which remains tantalizing and has been the subject of much recent work.  A striking property of the entropy formula~(\ref{bhe}) is that it is proportional to the black hole's area, not its volume.  For ordinary systems the number of atoms goes as the volume.  Here it is as though they are spread out on the horizon of the black hole, with a density of order one per Planck area.  Moreover, in any gravitating system, most states are black holes (if one tries to excite a lot of degrees of freedom, they will gravitationally collapse).  This suggests the {\it holographic principle,} that in quantum gravity the fundamental degrees of freedom for any region live on its surface, not in its volume~\cite{'tHooft:1993gx,Susskind:1994vu}.

\subsubsection{The information paradox}

The large number of black hole microstates reflects the fact that the black hole might be formed in many different ways, by collapsing matter from many possible initial quantum states.  Once this matter is behind the event horizon, causality prevents it from influencing anything outside the black hole.  In particular, the state of the Hawking radiation cannot depend on the state of the infalling matter.  But this implies irreversibility: after the black hole evaporates so that only the Hawking radiation remains, the final state of the system is essentially independent of the initial state~\cite{Hawking:1976ra}.  In this sense, information is lost.

This is inconsistent with normal quantum mechanical time evolution,
\be
i \partial_t \psi = H \psi \,.
\ee
This Schr\"odinger form holds not just in quantum mechanics, but also in quantum field theory and in string perturbation theory.  Such a differential equation, with a self-adjoint $H$, can be integrated backwards in time as easily as forward, and so the final state determines the initial state uniquely.  Hawking therefore proposed a more general evolution law~\cite{Hawking:1976ra}, the dollar matrix.

This is a true paradox.  It allows a number of alternatives, all of them seemingly unpalatable: (a) information is lost in the way that Hawking argued;
(b) information is carried away by the outgoing Hawking radiation; (c) the black hole does not evaporate completely, but ends in a Planck-sized remnant with an enormous number of internal states.  We will not review here all of the arguments in various directions, but will list a major criticism of each.  On (a), it is argued that dollar matrix evolution does not satisfy Noether's theorem, and energy conservation is violated in a radical way~\cite{Banks:1983by}.  On (b), information needs to travel faster than the speed of light.   This is not just the apparent nonlocality of the EPR effect, as its effect would be seen in actual measurements of the Hawking radiation. On (c), the large number of states is inconsistent with the entropy formula~(\ref{bhe}) which relates the number of states of a black hole to it size, and it may lead to uncontrolled virtual effects.

In fact, Matrix theory already tells us the answer~\cite{Banks:1996vh}.  The $S$-matrix defined by Matrix theory includes processes in which a black hole forms in a high energy collision and then evaporates.  Matrix theory is an ordinary quantum mechanical system, so Schr\"odinger evolution holds and information is not lost.  Moreover, the quantum mechanical spectrum does not have room for the states of a remnant.  So the answer must be (b).

This argument is now accepted, at the same level of belief as all the other dualities, but it was not immediately appreciated.  It was not initially clear that Matrix theory represents a full-fledged duality.  For one thing, the large-$N$ limit involved is somewhat unfamiliar, and the physical interpretation of the finite-$N$ theory is a bit exotic~\cite{Susskind:1997cw}.  So work on the information paradox continued, and was extremely fruitful.

\subsection{AdS/CFT duality}

We have noted that we can move back and forth between D-branes and black branes by dialing the coupling up and down.  Since D-branes seem to have a normal quantum mechanical description, this suggests that black branes should as well.  In order to make this more precise, after the agreement of entropies was found the dynamical properties of D- and black branes were compared, i.e.\ the amplitudes for the branes to absorb and emit radiation.
A series of unexpected coincidences were found, where very different calculations gave the same answer --- an operator matrix element on the D-brane side, and a curved spacetime propagator on the black brane side.  Primed by the previous examples, the reader will suspect a duality, and this was crystallized by Maldacena~\cite{Maldacena:1997re}.

Maldacena's insight was that one obtains a duality if one takes the low energy limit of the D/black brane.  As we have noted earlier, the low energy limit of the physics on a D-brane is a QFT, supersymmetric Yang-Mills theory.  The low energy limit of the black brane isolates the geometry near the horizon, because the gravitational redshift blows up there.\footnote{On both sides there are also massless gravitons moving in the bulk, but these decouple because they are propagating in a larger number of noncompact dimensions.}  The case that has received the most attention is 3-branes.  On the gauge side, one gets the ${\cal N}=4$ theory discussed earlier.  On the black side, the near-horizon geometry is $AdS_5 \times S^5$.  The claim is that the $d=4$, ${\cal N}=4$ $SU(N)$ Yang-Mills theory is dual to the IIB string on $AdS_5 \times S^5$. (The remaining $U(1)$ from $U(N)$ decouples.)    Here $AdS$ denotes anti-de Sitter spacetime, a solution to Einstein's equation that once was rather exotic, but whose special properties are essential to the duality.

An immediate check is that the symmetries match.  The $AdS_5 \times S^5$ spacetime has the geometrical symmetry $SO(4,2) \times SO(6)$.  In the QFT, the $SO(4,2)$ emerges from the Poincar'e symmetry plus conformal symmetry, the latter being a special property of the ${\cal N}=4$ theory, and the $SO(6)$ is a global symmetry of this theory.  The duality means that there is a 1-1 correspondence in the spectrum, with equality of observables.  The simplest observables in a QFT are expectation values of products of local operators.  In the string theory, these are dual to perturbations of the boundary conditions on $AdS_5$~\cite{GKP,W}.  In quantum gravitational theories, it is difficult to find simple observables --- because of the coordinate invariance and spacetime dynamics, one must specify where the observable is by including effectively a physical system of clocks and rods.  But anti-de Sitter space is rather special, in that it has a boundary where the quantum fluctuations go to zero, and observables there (such as perturbations of the boundary conditions) can be simple.

There are two parameters on each side of the duality.  In the Yang-Mills theory there is the coupling $g_{\rm YM}^2$ and the rank $N$ of the gauge theory.  In the string theory there is the string coupling $g_{\rm s}$ and the number $N$ of units of RR flux on the $S^5$.  The $N$'s are the same, and the couplings are related by $g_{\rm YM}^2 = 4\pi g_{\rm s}$.  It is useful to relate these parameters to the three characteristic length scales in the string theory description, the Planck length $l_{\rm P}$, the string length $l_{\rm s}$, and the curvature radius $l_{\rm AdS}$ of the $AdS_5$ and $S^5$:
\be
\frac{l_{\rm AdS}}{l_{\rm P}} \sim N^{1/4} \,,  \quad \frac{l_{\rm AdS}}{l_{\rm s}} \sim ( g_{\rm s}N)^{1/4} \,.
\ee
In order for a classical gravitational description of the AdS spacetime to be valid, we need both of these ratios to be large, so both the rank $N$ of the gauge group and the combination $\lambda =  g_{\rm YM}^2 N  \sim g_{\rm s} N$ must be large.  It was pointed out by 't Hooft~\cite{'tHooft:1973jz} that $\lambda$ is the parameter that controls the validity of the QFT perturbation theory: the dominant graphs at each loop order have one power of $N$ for each power of $g_{\rm YM}^2$.  So for large $N$, if $\lambda \ll 1$ then the weakly coupled QFT is the good description, and if $\lambda \gg 1$ then the string theory in AdS is the good description.

As an aside, the reader might note that we have proposed two duals for strongly coupled $d=4$, ${\cal N}=4$ gauge theory --- here the $AdS_5 \times S^5$ IIB theory, while in section 2.5 the original gauge theory but in magnetic variables.  The point is that there are three descriptions, with the magnetic one holding at very large $\lambda$.  The original QFT is good if $g_{\rm YM}^2 \ll 1/N$, so the dual QFT is good if $1/g_{\rm YM}^2 \ll 1/N$ meaning that $g_{\rm YM}^2 \gg N$.  In between, $1/N \ll g_{\rm YM}^2 \ll N$, the AdS string theory is the good description.

't Hooft also noted that the graphs that dominate at each order have a special structure.  They are planar, meaning that they can be drawn on a two-dimensional surface without crossing of lines.  This is suggestive of a string world-sheet, and 't Hooft proposed that the large-$N$ theory might be rewritten as a string theory~\cite{'tHooft:1973jz}.  This idea was tantalizing; see for example Ref.~\cite{Polyakov:1987ez} for further exploration of this.  One surprise about the final form of the duality is that it involves the exact same string theory that describes quantum gravity, but that the dual spacetime $AdS_5 \times S^5$ is very different from that in which the gauge theory lives.  The unusual properties of $AdS_5$, in particular its gravitational warping, explain why there is no graviton in the 4-dimensional spectrum.  't Hooft's connection with planar gauge theory is related to the existence of strings in the dual theory.

The QFT also contains the duals of black holes: they are thermal equilibrium states of high energy.  There is a phase transition in the QFT~\cite{Witten:1998zw}.  At low temperature the entropy is of order $N^0$.  This correspond to a gas of gravitons.  At high temperature the entropy is of order $N^2$.  Tracing through the parameters, this matches the entropy formula~(\ref{bhe}), generalized to ten dimensions.  In particular it reproduces the distinctive factor of $1/\hbar$, which is special to the black hole entropy.  

This is consistent with the intuition that black holes should behave like ordinary thermal objects: they are literally dual to a thermal state.  Further, AdS/CFT, like Matrix theory, can be used to argue for conservation of information.  One can consider a black hole forming and evaporating within an AdS box, and this has a QFT description in terms of ordinary quantum evolution.

\subsection{Gauge/gravity duality}

If we apply the same reasoning to D$p$-branes for $p \neq 3$, one again obtains a duality~\cite{Itzhaki:1998dd}.  The gauge theory is no longer conformally invariant, and the gravitational space has less symmetry as well.  Thus, the label AdS/CFT used for the D3 and other early examples is too special, and gauge/gravity duality is the more general term (though this still may not be general enough, perhaps QFT/gravity would be more encompassing). 

Note the analogy between the commutator-squared potential ${\rm Tr} ( [X^i, X^j] [X^i, X^j] )$ that was the key to Matrix theory and the quartic interaction ${\rm Tr} (  [ A_\mu ,  A_\nu]  [ A_\mu ,  A_\nu])$ in Yang-Mills theory.  This is more than a coincidence.  Matrix theory is the quantum mechanics of D0-branes, and $AdS_5 \times S^5$ is the quantum mechanics of D3-branes.  These D-branes are related by $T$-duality, so Matrix theory and AdS/CFT are both part of a larger web of gauge/gravity dualities.

Like Matrix theory, AdS/CFT provides a nonperturbative definition of quantum gravity and string theory.
Recalling the discussion of observables, the spacetime of the QFT is identified with the boundary of anti-de Sitter space.  Thus, AdS/CFT is a precise realization of the holographic principle:   the nonperturbative variables live on the boundary of the space of interest.

It is sometimes suggested that AdS/CFT might only be an approximate duality; see for example~\cite{Gary:2011kk}.  At one level it is difficult to see how this could be.  The CFT is a complete and well-defined quantum field theory.  It contains states that can be identified with gravitons, strings, and black holes in AdS spacetime, and many of the detailed properties agree with those expected for these particles.  What quantum theory could be so similar to quantum gravity (or, more precisely, the IIB string) without actually being quantum gravity?

In the D3 duality, the QFT is a 3+1 dimensional gauge theory.  These of course are interesting for many other reasons, and gauge/gravity duality is a new tool for understanding them.  When the 't Hooft parameter is large, we cannot use perturbation theory, but now we can use the dual gravitational description to calculate.  

QCD itself, the theory that we would like to solve, will not have any simple dual.  The problem is that the coupling is large only in a narrow range around the confinement scale.  At higher energies it is weak because of asymptotic freedom, and at lower energies there are no degrees of freedom.  Thus, there is not much room for a duality to operate.  By contrast, the coupling does not run in the ${\cal N}=4$ theory, so the coupling can be strong at all scales.  One can deform the conformal ${\cal N}=4$ theory to break some of its supersymmetry, and if enough is broken the theory would be expected to confine. Indeed, this is what one finds on the gravitational side.  One can even take a limit that should give true QCD, but it is complicated and will involve spacetimes of large curvature.

Surprisingly, even the conformal AdS/CFT duality seems to capture some features of the strongly interacting states that are produced at heavy ion accelerators~\cite{Kovtun:2004de}, and which are difficult to understand from any other point of view.  Gauge/gravity duality is also used to model strongly coupled phases in condensed matter systems~\cite{Hartnoll:2009sz}.  As with QCD, the theories that have sharp duals are always idealizations, and one must work to distinguish universal properties from artifacts of the model, but these dualities allow the study of a range of phemenomena in a novel way, orthogonal to standard methods.

Gauge/gravity duality, starting from the QFT side, is an example of emergent gravity.  In \S2 we discussed emergent gauge theory, so emergent gravity may not be such a surprise.  However, it took much longer to realize.  A celebrated no-go theorem~\cite{Weinberg:1980kq} suggested that it would be impossible.  The essence of this theorem is a remark we made earlier, about the paucity of observables in quantum gravity: QFT's without gravity have many more observables, so there was a mismatch.   As with other powerful but ultimately falsified no-go theorems such as~\cite{Coleman:1967ad},  it is evaded but only due to rich and unexpected new ideas.  For emergent gravity via gauge/gravity duality, it was necessary that new dimensions emerge as well.  The mismatch of observables is resolved as we have discussed by identifying the QFT with the boundary of the gravitational space.  And along with gravity and spacetime dimensions, strings, or more generally M theory, emerge as well.

\sect{Discussion}

\subsection{Open questions}

Duality has given us a vastly larger and more precise picture of string/M theory.  Perturbation theory covered only small neighborhoods of the cusps of the duality diagram.  String/string dualities let us see the whole picture.  Gauge/gravity dualities then provided exact descriptions of special regions, and joined QFT into the web.

What we are still missing is a global definition of M theory.  The dual constructions are limited to theories in special spaces such as AdS, which have visible boundaries.  Extending this to cosmological spacetimes such as our own is a large step.  One attempt is dS/CFT~\cite{Strominger:2001pn,Witten:2001kn}, in which the future infinity of de Sitter replaces the boundary of AdS, and other frameworks are being explored as well~\cite{Alishahiha:2004md}, but no clear success has emerged.  Related to this is the problem of extending the holographic principle in a precise way to situations where there are not special boundaries.

Black hole quantum mechanics has been a fruitful laboratory, and we may have more to learn from it.   Gauge/gravity duality tells us that information is carried away by the Hawking radiation, which seems tantamount to traveling faster than light, but the duality does not tell us how this works.  The recent black hole firewall paradox~\cite{Almheiri:2012rt} sharpens the issue.  It had been widely believed that no single observer would see any violation of the ordinary laws of physics, but this seems to lead to a contradiction: if all is normal for observers outside the black hole, then an infalling observer sees something very different from a smooth event horizon.  

Most attempts to avoid this conclusion relax the rules of quantum mechanics --- not in the way that Hawking proposed, which would be visible to an external observer, but in the description of the infalling observer.  At this point the subject is much like the original information paradox before gauge/gravity duality.  Theorists are trying out different scenarios, in which one or another assumption is relaxed, but a convincing theory has yet to emerge.  Gauge/gravity duality itself has been rather impotent here.  It gives a precise, and as far as we can tell complete, description of measurements that can be made by an observer at infinity.  It had been widely believed that one could infer from the QFT a description of the black hole interior as well, but the firewall argument has challenged this.  It seems that the dual QFT does not tell us whether the firewall is there in the gravitational theory, or perhaps  we do not yet have the dictionary needed to interpret it.

We have repeatedly discussed emergent space, but what of emergent time?  In all gauge-gravity duals (except the poorly understood IKKT model~\cite{Ishibashi:1996xs}), time is already present on the QFT side.  Even so, there is some measure of emergent time: due to relativity, there is not a direct identification between time in the QFT and time in the gravitational bulk.  But the understanding of this is far from precise.  An interesting  recent direction has been the study of holographic entanglement entropy~\cite{Nishioka:2009un}, relating spacetime geometry to entanglement in the dual \mbox{QFT}.  It is not yet clear what the ultimate lesson will be, but this brings together ideas from several fields, and is being pursued vigorously.

On the list of unsolved problems, one should mention the understanding of string vacua like ours, with broken supersymmetry.  There are some constructions of these, but they rest on multiple approximations and no exact theory.  For example, the value of the cosmological constant in such a vacuum should be calculable algorithmically to essentially perfect accuracy (even if this is not a practical thing to do).  In a complete theory there must be a prescription for this.

The problems of the black hole interior, and of nonsupersymmetric vacua, both point to limits in our current understanding.  The issue may be that we do not have an independent nonperturbative construction of string/M theory.  Rather, at present we have a nonperturbative construction of the dual QFT, and various approximations on the string/M side.  Perhaps, with improved technology, this will be enough, but it seems that we are still missing some big idea.  Duality has revealed much of the fascinating mathematical/physics structure known as M theory, but its full form is still for us to discover.

\subsection{Fundamentals}

One should always be a bit agnostic about what is fundamental in the current understanding of the laws of nature.  Future technical developments may completely change the landscape, and what looks central today might be a sideshow tomorrow.

{\it Symmetry} is a cautionary example.  In the 1960's and 1970's, symmetry reigned, from the global $SU(3)$ of Gell-Mann and Ne'eman to the local $SU(3)\times SU(2) \times U(1)$ of the Standard Model, and beyond to grand unification and supersymmetry.  In the context of the Standard Model, global $SU(3)$ is now seen as an accident, which appears in an approximate way because the scale of the quark masses happens to be somewhat less than the scale of quark confinement.  Indeed, when this ratio is reversed a different approximate global symmetry appears instead~\cite{Isgur:1989vq}.  The structure of the Standard Model is determined entirely by its gauge symmetries.  It does have one global symmetry, $B-L$ (baryon number minus lepton number), but this again appears to be an accident: given the gauge symmetries and particle content, there are no renormalizable terms that could violate $B-L$.\footnote{Baryon number and lepton number separately are not symmetries of the Standard Model due to anomalies.  Anomaly mediated $B$ and $L$ violation is believed to have been substantial in the early universe.}  This point of view would suggest that a small violation, from virtual high energy effects, would eventually be seen, and the neutrino masses may be evidence for this.  From more theoretical points of view, string theory appears to allow no exact global symmetries~\cite{Banks:1988yz}, and in any theory of quantum gravity virtual black holes might be expected to violate all global symmetries~\cite{Hawking:1974sw}.

Moreover, as we have already discussed in \S2, local (gauge) symmetries have been demoted as well, with the discovery of many and varied systems in which they emerge essentially from nowhere.  It seems that local symmetry is common, not because it is a basic principle, but because when it does emerge it is rather robust: small perturbations generally do not destroy it.  Indeed, it has long been realized that local symmetry it is `not really a symmetry,' in that it acts trivially on all physical states.    The latest nail in this coffin is gauge/gravity duality, in which general coordinate invariance emerges as well.

This leaves us in the rather disturbing position that no symmetry, global or local, should be fundamental (and we might include here even Poincar\'e invariance and supersymmetry).  Susskind has made a distinction between the mathematics needed to write down the equations describing nature, and the mathematics needed to solve those equations~\cite{Suss}.  Perhaps symmetry belongs only to the latter.\footnote{As an example, the $SO(4,2) \times SO(6)$ symmetry of the prototypical gauge/gravity duality is a symmetry of the $AdS_5 \times S^5$ solution to the bulk field equations, but there are many other solutions of lesser or no symmetry.  On the CFT side this $SO(4,2) \times SO(6)$ is a symmetry of the Lagrangian.  In QFT, different Lagrangians would be regarded as defining different theories, but here there is a larger structure where the CFT Lagrangian is contingent on the solution.  One might think of it loosely as an `effective Lagrangian.'}   Or perhaps this is taking things too far.

{\it Stringiness} is another example.  After the early successes of string theory, a tempting direction was to seek a nonperturbative formulation of the theory in terms of the quantum mechanics of strings.  But work in this direction, such as abstract conformal field theory, and string field theory, seem not to have been so fruitful.  Rather, strings have been a bridge to the discovery of M theory, in which they are found to emerge  in some classical limits but not in others.

What then are we to make of the role of {\it duality} in recent developments?  Let me take a pragmatic point of view.  When we deal with a large quantum system, we have a lot of knobs to turn: couplings can be weak or strong, spaces can be large or small.  When parameters are taken to extreme values, it might be that the physics becomes chaotic or turbulent and has no simple description.  But in a large number of physical systems, it is possible to find some simplification that makes such limits tractable.  And when the system is constrained by quantum mechanics, and Poincar\'e invariance, and possibly supersymmetry as well, then limits often have a classical interpretation.  Moreover, it seems that these principles are so constraining that  in many cases it is possible to reconstruct the whole quantum theory from such a limit.

The existence of dualities points to a great unity in the structure of theoretical physics.  String-string dualities imply that there is a unique string/M theory.  {\it Unique} here means that there are not even adjustable coupling constants, as in quantum field theory; rather, the apparent parameters are properties of the particular state (vacuum) of the system.  Gauge/gravity dualities further imply that many QFT's can also be interpreted as vacua of string theory.  As we have noted, a necessary condition for a QFT to have an interpretation in terms of a geometrical spacetime is that the number of fields be large.\footnote{There is some evidence that this condition, plus the condition that the QFT be strongly coupled in a certain precise technical sense (large anomalous dimensions), actually constitute sufficient conditions~\cite{Heemskerk:2009pn}.}  However, there is no sharp cutoff on the number of fields, so QFT's with small numbers of fields should correspond to string theories in highly curved spaces.  In this sense it may be that every QFT can be understood as a vacuum state of string/M theory.

When we identify the different duality frames with different fundamental descriptions, we are essentially saying that what is fundamental is what we see in the classical limit.  This is strange, in a world that is intrinsically quantum mechanical  (and what of theories that have no classical limit?).  But our universal tool for constructing relativistic quantum field theories, the path integral, begins with a sum over classical histories.  Somehow we do not have a fully quantum point of view, and this may be connected with our inability to derive the interesting dualities.  We want to know what is the theory in the middle of the duality diagram.

Our understanding of what is fundamental has taken many twists and turns, some of which I have tried to describe.  I have also tried to describe some of the methods and reasoning that have brought us to our current level of understanding.  The immediate questions are, how do we go beyond this, and which of the current threads will be the productive ones?

\section{Acknowledgments}

I am grateful to many colleagues over the years, beginning with my advisor Stanley Mandelstam, for insights into duality.  I would like to thank Stanley Deser, Dieter Luest, Fernando Quevedo, Eliezer Rabinovici, Lo\"ic Turban, and especially Bill Zajc for their comments on the manuscript. This work was supported in part by NSF grants PHY11-25915 and PHY13-16748.

\end{document}